\documentclass[letterpaper]{article} %

\usepackage{aaai24}
\usepackage{times}  %
\usepackage{helvet}  %
\usepackage{courier}  %
\usepackage[hyphens]{url}  %
\usepackage{graphicx} %
\usepackage{natbib}  %
\usepackage{caption} %
\usepackage{aas_macros}
\usepackage{amsmath}
\usepackage{amsthm}
\usepackage{amssymb}
\usepackage{booktabs}
\usepackage{chemformula}
\usepackage{cleveref}
\usepackage{csquotes}
\usepackage{derivative}
\usepackage{microtype}
\usepackage{multirow}
\usepackage{nicefrac}
\usepackage{siunitx}
\usepackage{subcaption}
\usepackage{tabularx}
\usepackage[flushleft]{threeparttable}

\usepackage{standalone}
\usepackage{tikz}
\usetikzlibrary{shapes.geometric}
\usepackage{pgfplots}
\pgfplotsset{compat=1.18}
\pgfmathsetseed{42}

\newcommand{\given}{\ensuremath{\,|\,}}

\usetikzlibrary{patterns}
\usepackage[precision=2, unit=mm]{lengthconvert}

\usepackage{colortbl}
\usepackage{hhline}
\definecolor{low}{HTML}{081d58}
\definecolor{mid}{HTML}{40b5c4}
\definecolor{high}{HTML}{ffffd9}
\newcommand*{\MinNumber}{0.0}%
\newcommand*{\MidNumber}{90.0} %
\newcommand*{\MaxNumber}{180.0}%
\newcommand{\cc}[1]{%
    \pgfmathsetmacro{\foo}{#1 / \MaxNumber}
    \ifdim \foo pt > 0.5 pt
        \pgfmathparse{max(min(100.0*(#1 - \MidNumber)/(\MaxNumber-\MidNumber),100.0),0.00)}%
        \xdef\tempa{\pgfmathresult}%
        \cellcolor{low!\tempa!mid}%
        \textcolor{white}{#1}%
    \else
        \pgfmathparse{max(min(100.0*(\MidNumber - #1)/(\MidNumber-\MinNumber),100.0),0.00)}%
        \xdef\tempa{\pgfmathresult}%
        \textcolor{black}{#1}%
        \cellcolor{high!\tempa!mid}%
    \fi
}

\newcolumntype{C}{>{\centering\arraybackslash}X}
\newcolumntype{R}[2]{%
    >{\adjustbox{angle=#1,lap=\width-(#2)}\bgroup}%
    l%
    <{\egroup}%
}
\newcommand*\rot{\multicolumn{1}{R{45}{-0.5em}}}

\title{Inferring Atmospheric Properties of Exoplanets with\\ Flow Matching and Neural Importance Sampling}
\author{
    T.~D.~Gebhard\textsuperscript{1,2}, 
    J.~Wildberger\textsuperscript{1}, 
    M.~Dax\textsuperscript{1}, 
    D.~Angerhausen\textsuperscript{2}, 
    S.~P.~Quanz\textsuperscript{2,3,$\dagger$}, 
    B.~Schölkopf\textsuperscript{1,4,$\dagger$}
}
\affiliations{
    \footnotesize
    \textsuperscript{1}Max Planck Institute for Intelligent Systems, Max-Planck-Ring 4, 72076 Tübingen, Germany\\
    \textsuperscript{2}ETH Zurich, Institute for Particle Physics \& Astrophysics, Wolfgang-Pauli-Strasse 27, 8093 Zurich, Switzerland\\
    \textsuperscript{3}ETH Zurich, Department of Earth Sciences, Sonneggstrasse 5, 8092 Zurich, Switzerland\\
    \textsuperscript{4}ETH Zurich, Department of Computer Science, Universitätsstrasse 6, 8092 Zurich, Switzerland\\[1mm]
    \{tgebhard\,,\,jwildberger\,,\,mdax\,,\,bs\}@tue.mpg.de, \{dangerhau\,,\,sascha.quanz\}@phys.ethz.ch 
    \qquad \qquad 
    \textsuperscript{$\dagger$}joint supervision
}

\nocopyright

\begin{document}

\maketitle

\begin{abstract}
    Atmospheric retrievals (AR) characterize exoplanets by estimating atmospheric parameters from observed light spectra, typically by framing the task as a Bayesian inference problem.
    However, traditional approaches such as nested sampling are computationally expensive, thus sparking an interest in solutions based on machine learning (ML).
    In this ongoing work, we first explore flow matching posterior estimation (FMPE) as a new ML-based method for AR and find that, in our case, it is more accurate than neural posterior estimation (NPE), but less accurate than nested sampling. 
    We then combine both FMPE and NPE with importance sampling, in which case both methods outperform nested sampling in terms of accuracy and simulation efficiency. 
    Going forward, our analysis suggests that simulation-based inference with likelihood-based importance sampling provides a framework for accurate and efficient AR that may become a valuable tool not only for the analysis of observational data from existing telescopes, but also for the development of new missions and instruments.
\end{abstract}

\section{Introduction}

\enquote{NASA Says Distant Exoplanet Could Have Rare Water Ocean} \citep{Luscombe_2023} --- Headlines like this have recently made it even into the mainstream news.
But how do we know what happens on (and above) the surface of planets outside our solar system?
In many cases, the answer is \emph{atmospheric retrievals} (AR), that is, \enquote{the inference of atmospheric properties of an exoplanet given an observed spectrum} \citep{Madhusudhan_2018}.
These properties include the abundances of chemical species (e.g., water or methane), the thermal structure, or the presence of clouds.
In practice, performing an AR usually means combining a simulator for the forward direction (i.e., parameters $\to$ spectrum) with a Bayesian inference technique such as nested sampling \citep{Skilling_2006,Ashton_2022} to compute a posterior distribution over the atmospheric parameters of interest.
Depending on the complexity of the simulator, the number of parameters, the spectral resolution of the observed data, and other factors, this can become very computationally expensive:
A single AR can easily require on the order of tens of thousands of CPU hours, often resulting in wall times of days to weeks.

Reducing this computational burden has already attracted the attention of the machine learning (ML) community, including even a competition at NeurIPS~2022 \citep{Changeat_2022}.
Previously proposed ML approaches to the problem of AR include the usage of GANs \citep{Zingales_2018}, random forests \citep{MarquezNeila_2018, Fisher_2020}, Monte Carlo dropout \citep{Soboczenski_2018}, Bayesian neural networks \citep{Cobb_2019}, various deep learning architectures \citep{Yip_2021,ArdevolMartinez_2022,Giobergia_2023,Unlu_2023}, variational inference \citep{Yip_2022}, and neural posterior estimation (NPE) using discrete normalizing flows \cite{Vasist_2023}.
NPE was also used by the winning entry to the 2023 edition of the ARIEL data challenge \citep{Aubin_2023}.
Finally, a related but somewhat orthogonal direction are the approaches by \citet{Himes_2022} and \citet{Hendrix_2023}, who do not predict a posterior directly, but instead speed up ARs by replacing the computationally expensive simulator with a learned emulator.

In this workshop paper, we first introduce another ML approach to AR: flow matching posterior estimation (FMPE) using continuous normalizing flows. 
Focusing on one specific case study from the literature, we then compare FMPE to both a nested sampling approach and neural posterior estimation (NPE) as introduced by \citet{Vasist_2023}.
Finally, we combine both FMPE and NPE with neural importance sampling and show that this improves the results significantly.\looseness=-1

\section{Method}

\begin{figure*}[t]
    \centering
    \begin{subfigure}[b]{8.4cm}
         \centering
         \includegraphics{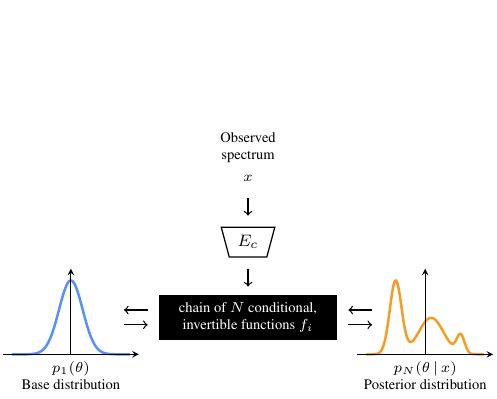}
         \caption{NPE}
         \label{fig:npe}
     \end{subfigure}
     \hfill
     \begin{subfigure}[b]{8.4cm}
         \centering
         \includegraphics{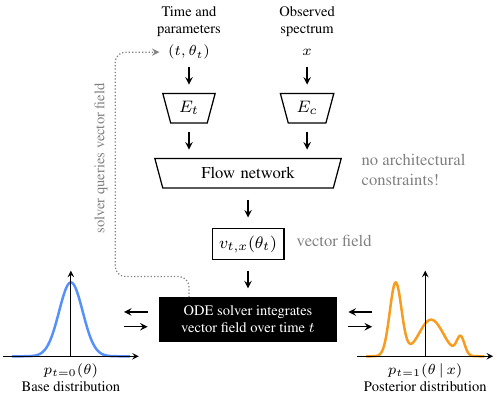}
         \caption{FMPE}
         \label{fig:fmpe}
     \end{subfigure}
     \caption{
        Schematic comparison between neural posterior estimation (NPE) and flow matching posterior estimation (FMPE).
     }
     \label{fig:npe-fm}
\end{figure*}

We briefly recapitulate NPE, which will serve as another baseline besides nested sampling, and then introduce the FMPE method as well as the idea of neural importance sampling.
A schematic comparison of NPE and FMPE is found in \cref{fig:npe-fm}.

\paragraph{NPE with normalizing flows}
NPE \citep{Papamakarios_2016} is a technique for simulation-based inference (SBI; \citealt{Cranmer_2020}) that trains a density estimator $q(\theta \given x)$ to approximate the posterior $p(\theta \given x)$ by minimizing the following loss:
\begin{align}
    \mathcal{L}_\textrm{NPE} = -\mathbb{E}_{\theta \sim \pi(\theta)} \  \mathbb{E}_{x \sim p(x \given \theta)} \log q(\theta  \given  x) \,.
    \label{eq:loss-npe}
\end{align}
Here, $\pi(\theta)$ denotes the prior and sampling from the likelihood corresponds to a call of the forward simulator. 
Once trained, $q(\theta \given x)$ serves as a surrogate for the posterior, enabling cheap sampling and density evaluation. 
The density $q(\theta \given x)$ is often parameterized with a conditional discrete normalizing flow \citep[DNF;][]{Tabak_2010,Rezende_2015}. 
DNFs construct the distribution
\begin{align}
    q(\theta \given x) 
    = p_0\left(\psi_x^{-1}(\theta)\right) \cdot \det \left| \frac{\odif{\psi_x^{-1}(\theta)}}{\odif{\theta}} \right|
    \label{eq:DNF}
\end{align}
by applying a chain of invertible functions $\psi_x: \mathbb{R}^n \to \mathbb{R}^n$, $\psi_x = f_N \circ \cdots \circ f_2 \circ f_1$ to a simple base distribution $p_0$ (e.g., an $n$-dimensional Gaussian). 
To ensure invertibility and efficient computation of the Jacobian in \cref{eq:DNF}, most DNFs impose strong constraints on the architecture.

\paragraph{FMPE}
Continuous normalizing flows \citep[CNF;][]{Chen_2018a} also transform a simple base distribution to a more complex one, but describe this transformation continuously. 
Specifically, the sample trajectories $\psi_{t,x}$ are parameterized in terms of a \enquote{time} parameter $t \in [0, 1]$, and a vector field $v: [0, 1] \times \mathbb{R}^{m+n} \to \mathbb{R}^{n}$, where $m = \dim(x)$ and $n = \dim(\theta)$, defined by the ordinary differential equation (ODE):
\begin{align}
    \odv{}{t} \psi_{t, x}(\theta) = v_{t, x}\left(\psi_{t, x}(\theta) \right) \,,
    \quad
    \psi_{0, x}(\theta) = \theta \,.
\end{align}
Conversion between the base ($t=0$) and target ($t=1$) distribution is then achieved by integration,
\begin{align}
    q_1(\theta \given x)
    = q_0(\theta) \cdot \exp\left\{ -\int_0^1 \operatorname{div} v_{t,x}(\theta_t) \odif{t} \right\} \,.
    \label{eq:cnf-density}
\end{align}
CNFs offer great flexibility, as they are parameterized by unconstrained vector fields and thus do not impose architectural constraints. However, likelihood maximization can be prohibitively expensive due to the cost of the ODE integration.

Flow matching \citep{Lipman_2022} provides an alternative training objective that directly regresses $v$ onto a target vector field $u$ by minimizing $\mathbb{E}\left[ || v - u ||^2 \right]$. 
It has been shown that the target can be carefully designed as a sample-conditional vector field $u_t(\theta \given \theta_{t=1})$, resulting in a tractable and efficient training objective. 
In the context of SBI, flow matching has been explored by \citet{Dax_2023b}, and the resulting method (flow matching posterior estimation; FMPE) maintains the desirable properties of NPE (expressiveness of the distribution, tractable density, simulation-based training) without requiring constrained neural architectures. 
In comparison to NPE, FMPE training is typically faster (due to simpler architectures) and inference is slower (due to the ODE integration).

\paragraph{Neural importance sampling}
In practice, both NPE and FMPE results may deviate from the exact posterior due to insufficient training data or network capacity, or when confronted with out-of-distribution (OOD) data. 
Further, it is typically difficult to assess whether an inferred posterior is accurate without comparing to results from another (trusted) inference method, which may not always be available: While nested sampling has been hailed as the gold standard for AR, some implementations have already been found to produce overly confident results \citep[e.g.,][]{ArdevolMartinez_2022}.

One way to address these challenges is to combine SBI methods with likelihood-based importance sampling \citep{Dax_2023a}. 
In this case, the inferred estimate $q(\theta \given x)$ is used as a proposal distribution for importance sampling \citep[IS;][]{Kloek_1978} by attaching importance weights\looseness=-1
\begin{align}
    w_i 
    = p(\theta_i \given x) \cdot p(\theta_i) / q(\theta_i \given x) \,
\end{align}
to each sample $\theta_i\sim q(\theta \given x)$. This transforms $N$ samples from $q(\theta \given x)$ into weighted samples from the true posterior $p(\theta \given x)$. 
\citet{Dax_2023a} showed that this results in asymptotic recovery of the exact posterior, and that failures (e.g., due to OOD data) are marked by a low sampling efficiency 
\begin{align}
    \epsilon
    = \frac{1}{N} \cdot \left. \left(\sum_{i=1}^{N} w_i\right)^2 \middle/ \sum_{i=1}^{N} w_i^2 \right. \,,
    \quad
    \epsilon \in [0, 1] \,.
\end{align}
The sampling efficiency also provides a direct performance measure: the better $q(\theta \given x)$ matches $p(\theta \given x)$, the higher the sampling efficiency. 
In practice, however, $\epsilon$ is susceptible to slight mismatches in even just one dimension of~$\theta$, resulting in a high variance of the~$w_i$ and thus low~$\epsilon$.

\section{Experiments and results}

We empirically evaluate both FMPE and NPE on the benchmark retrieval case from \citet{Vasist_2023}, which is based on a study of the planet HR~8799~e by \citet{Molliere_2020}.

\paragraph{Simulator}
We use the simulation code from \citet{Vasist_2023}, which itself is based on the \texttt{petitRADTRANS} simulator (\citealt{Molliere_2019}; we used v2.6.7). 
This maps a $\dim(\theta) = 16$ dimensional parameter space (see \cref{tab:atmospheric-parameters} for descriptions and priors) to simulated emission spectra for a gas giant-type planet (cf. \cref{fig:spectrum}). 
We work at a spectral resolution of $R = \Delta\lambda / \lambda = 1000$ (compared to $R = 400$ in \citealt{Vasist_2023}) with a wavelength range of \qtyrange{0.95}{2.45}{\micro\meter}, corresponding to $\dim(x) = 947$ bins. 
Following \citet{Vasist_2023}, we apply independent Gaussian noise with $\mu = 0, \sigma = 0.1257$ for each bin of the spectrum.

\paragraph{Nested sampling baseline}
We use \texttt{nautilus} \citep{Lange_2023} as a baseline, which implements importance nested sampling enhanced with deep learning. 
More conventional samplers such as \texttt{PyMultiNest} \citep{Buchner_2014} or \texttt{dynesty} \citep{Speagle_2020} did not converge after several weeks. 
Using the Gaussian likelihood implied by our simulator, we run with \num{10000} live points, \qty{0.1}{\percent} remaining live points as convergence criterion, a target effective sample size of \num{50000}, and default values for all other settings.

\paragraph{NPE and FMPE models}
We train both an NPE and an FMPE model.
The configurations reported here are the respective best ones from our preliminary experiments.

As illustrated in \cref{fig:npe-fm}, the NPE model consists of two parts: 
(1)~a context embedding network $E_c$ for the flux values $x$ in the form of \num{15} residual blocks of decreasing size (from \num{4096} to \num{256}) that outputs a representation $z \in \mathbb{R}^{256}$, and 
(2)~a neural spline flow \citep{Durkan_2019} with 20 piecewise rational quadratic coupling transforms (hidden size \num{1024}, \num{4} blocks, \num{16} bins) which are conditioned on~$z$.
For FMPE, the model has three parts: 
(1)~a context embedding network $E_c$ for the flux $x$ with two residual blocks,
(2)~a residual network $E_t$ with positional encodings applied to $t$, mapping $t$ and $\theta_t$ to a \num{512}-dimensional embedding, and 
(3)~the flow network with \num{40} residual blocks of decreasing size (from \num{8192} to \num{16}) that receives the embedded spectrum and $(t, \theta_t)$-tuple and predicts the vector field $v_{t, x}(\theta_t)$.
The total number of trainable parameters is 318\,M for NPE and 501\,M for FMPE.

We train our models for up to \num{1000} epochs on a dataset of 16.8\,M simulations with batch size \num{16384}, using the AdamW optimizer \citep{Loshchilov_2017} with a Reduce\-LR\-OnPlateau scheduler (initial learning rate $10^{-4}$, patience \num{30} epochs, factor \num{0.5}), early stopping (patience \num{100} epochs), and gradient clipping ($L_2$ norm $\leq$ \num{1.0}).
Both models use dropout; for FMPE, batch normalization also proved beneficial.
We use a 98\%\,/\,2\% split between training and validation, and only store the models that achieve the lowest validation loss.
Following standard ML practices, the atmospheric parameters $\theta$ are standardized by subtracting the mean and dividing by the standard deviation. 
Like in \citet{Vasist_2023}, the flux values $x$ are rescaled as $x \mapsto x / (1 + |x / 100|)$.
For FMPE, we train with automatic mixed precision (AMP) as it speeds up training significantly, while for NPE, we find that AMP has almost no effect.
On a single NVIDIA H100 GPU, training to convergence takes approximately \qty{54}{\hour} for the FMPE model (\num{747} epochs, \qty{261}{\second} per epoch), and about \qty{148}{\hour} (\num{1000} epochs, \qty{533}{\second} per epoch) for NPE.

\begin{figure}
    \centering
    \includegraphics{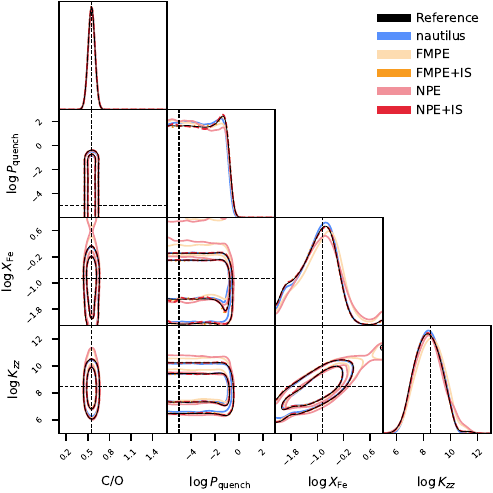}
    \caption{
        Comparison of the posterior estimates for four parameters; see \cref{fig:cornerplot} in the Appendix for the full version.
    }
    \label{fig:small-cornerplot}
\end{figure}

\begin{table*}[t]
    \centering
    \caption{
        JS divergence (in mnat) between the marginals of our reference posterior and the different methods (lower is better).
    }
    \label{tab:jsd-values}
    \scriptsize
    \begin{tabularx}{\linewidth}{
        @{}
        l*{16}{C}
        !{\color{white}\vrule width 2pt}
        c
    }
        Method & 
        \rot{C/O} & 
        \rot{[Fe/H]} & 
        \rot{log~P\textsubscript{quench}} & 
        \rot{log~X\textsubscript{Fe}} & 
        \rot{log~X\textsubscript{MgSiO\textsubscript{3}}} & 
        \rot{f\textsubscript{sed}} & 
        \rot{log~K\textsubscript{zz}} & 
        \rot{$\sigma_g$} & 
        \rot{log~$g$} & 
        \rot{R\textsubscript{P}} & 
        \rot{T\textsubscript{0}} & 
        \rot{$\frac{\text{T}_\text{3}}{\text{T}_\text{connect}}$} & 
        \rot{$\nicefrac{\text{T}_\text{2}}{\text{T}_\text{3}}$} & 
        \rot{$\nicefrac{\text{T}_\text{1}}{\text{T}_\text{2}}$} & 
        \rot{$\alpha$} & 
        \rot{$\frac{\text{log}~\delta}{\alpha}$} &
        Mean \\
        \midrule
        nautilus & \cc{1.7} & \cc{1.5} & \cc{37.9} & \cc{53.3} & \cc{20.3} & \cc{20.8} & \cc{56.7} & \cc{9.6} & \cc{7.1} & \cc{10.5} & \cc{12.5} & \cc{4.3} & \textbf{\cc{5.6}}  & \cc{4.3} & \cc{7.3} & \cc{9.1} & \cc{16.4}
        \\
        \noalign{\color{white}\hrule height 2.5pt}%
        FMPE & \cc{6.6} & \cc{11.1} & \cc{20.0} & \cc{107.1} & \cc{106.1} & \cc{69.2} & \cc{86.0} & \cc{17.9} & \cc{34.4} & \cc{17.1} & \cc{44.3} & \cc{21.6} & \cc{15.4} & \cc{19.3} & \cc{51.2} & \cc{46.8} & \cc{42.1}
        \\
        FMPE+IS & \cc{0.6} & \textbf{\cc{0.4}}  & \textbf{\cc{3.2}}  & \cc{6.7} & \textbf{\cc{4.9}}  & \textbf{\cc{4.3}}  & \cc{8.1} & \textbf{\cc{4.6}}  & \textbf{\cc{0.7}}  & \cc{3.7} & \cc{1.9} & \textbf{\cc{3.7}}  & \cc{5.6} & \textbf{\cc{3.5}}  & \cc{5.4} & \cc{2.1} & \cc{3.7}
        \\
        \noalign{\color{white}\hrule height 2.5pt}%
        NPE & \cc{14.8} & \cc{29.1} & \cc{22.9} & \cc{114.7} & \cc{110.7} & \cc{160.0} & \cc{83.4} & \cc{13.3} & \cc{28.6} & \cc{25.4} & \cc{65.1} & \cc{33.2} & \cc{16.1} & \cc{11.1} & \cc{72.5} & \cc{59.0} & \cc{53.8}
        \\
        NPE+IS & \textbf{\cc{0.2}}  & \cc{0.4} & \cc{4.1} & \textbf{\cc{4.5}}  & \cc{8.7} & \cc{4.4} & \textbf{\cc{7.0}}  & \cc{8.0} & \cc{1.6} & \textbf{\cc{1.6}}  & \textbf{\cc{1.6}}  & \cc{5.6} & \cc{6.3} & \cc{5.7} & \textbf{\cc{4.8}}  & \textbf{\cc{1.8}}  & \cc{4.1}
    \end{tabularx}
\end{table*}

\paragraph{Reference posterior}
Thorough quantitative evaluation of inference results requires comparison to reference posteriors. 
Even nested sampling may produce inaccurate results, so we here combine all three methods to produce a reference posterior with the approach proposed in \citet{Dax_2023a}.
For our given benchmark spectrum, we first generate a large number of approximate posterior samples (7\,M samples, equally distributed between \texttt{nautilus}\,/\,NPE\,/\,FMPE).%
\footnote{
    Generation with NPE / FMPE is cheap; for \texttt{nautilus}, we use the (correlated) intermediate samples to save computational cost.
} 
Then, we train an unconditional DNF $q(\theta)$ to estimate the distribution of these samples using a maximum log-likelihood objective. 
Finally, we generate weighted posterior samples with importance sampling, $\theta_i\sim q(\theta)$, $w_i = \pi(\theta_i) \cdot p(x \given \theta_i) / q(\theta_i)$. 
With this method, we generate $n_\text{eff} =$ 616\,k samples ($\epsilon = \qty{6.2}{\percent}$), which represent our reference posterior. 
Importance sampling is asymptotically exact if the proposal covers the entire target. 
For us, this is the case if the initial 7\,M samples cover the posterior support because density recovery with the DNF is performed using a probability mass covering training objective; see the discussion in \citet{Dax_2023a} for details.

\paragraph{Evaluation}
We now compare all methods: \texttt{nautilus}, FMPE, FMPE-IS (i.e., FMPE augmented with importance sampling), NPE and NPE-IS. 
For each method, we generate 50\,k effective samples. 
With importance sampling, we find efficiencies of $\epsilon = \qty{13.0}{\percent}$ for FMPE-IS and $\epsilon = \qty{2.5}{\percent}$ for NPE-IS. 
Qualitative results are shown in \cref{fig:small-cornerplot} for selected and in \cref{fig:cornerplot} for all parameters.
For a quantitative evaluation, we compare each result to the reference in terms of the Jensen-Shannon divergence (JSD) between the 1D marginal distributions (\cref{tab:jsd-values}). 
As an additional accuracy measure, we report upper bounds on the linear optimal transport distance in the Appendix (\cref{fig:ot-results}), which also captures high-dimensional distributional differences.

First, we see that nested sampling clearly outperforms standard FMPE and NPE in terms of accuracy. 
However, when FMPE and NPE are augmented with IS, their performance improves by an order of magnitude and their accuracies even exceed the \texttt{nautilus} baseline.
The deviations between FMPE-IS, NPE-IS and the reference are negligibly small, which is expected as all three of these estimates are asymptotically exact.
In practice, we expect that even the deviation of nested sampling from the reference will be scientifically irrelevant, and that the main advantage of FMPE-IS and NPE-IS is not improved accuracy, but reduced computational cost, especially in an amortized setting (see below).
Further, we note that FMPE consistently produces slightly more accurate results than NPE. 
Future work needs to investigate if this only holds for our specific case, or if this is a general trend.
Lastly, we observe that both FMPE and NPE struggle with similar parameters (e.g., log\,X\textsubscript{Fe}), which could indicate that the main challenge for the model lies in the extraction of relevant information from the spectrum, and not in the density estimation.\looseness=-1

\paragraph{Runtime considerations}
Besides the training time (2 days for FMPE, 6 days for NPE), the total runtime of our methods is additionally composed of the time to generate the dataset and to do inference.
On a single core of an AMD EPYC 7662 CPU, simulating one spectrum (with random parameters drawn from the prior) takes about $3.2\,\pm\,0.7\,\rm{s}$, implying a total of about \num{15000} CPU hours that can be arbitrarily parallelized.
For example, with 16 AMD EPYC 7662 CPUs (with 64 cores each), simulating our training set would take about \qty{15}{\hour}.

At inference, sampling and evaluating the log-probabilities from the trained model is almost negligible in terms of computational cost compared to the \texttt{nautilus} baseline: 
On a single GPU, this takes about \qty{12}{\second} for NPE, and about \qty{12}{\minute} for FMPE (using a tolerance of $10^{-3}$ for the ODE solver).
However, this is also arbitrarily parallelizable.
For IS, we additionally have to consider the cost for simulating spectra:
Assuming a sampling efficiency of $\epsilon = 5\%$, generating 50\,k effective samples requires another 900 single core hours, or less than one hour when assuming 16 CPUs.
This means that if we can make proper use of the parallelization capabilities of NPE and FMPE, it seems plausible that we can beat the wall time of nested sampling even in a non-amortized setting (i.e., when running only a single retrieval), which in our case was 8.5 days for \texttt{nautilus}, and would be much higher for traditional samplers.
Of course, this computational advantage becomes much more significant once we consider multiple retrievals.
\Cref{tab:runtimes} summarizes the expected inference costs based on our experiments.

\begin{table}[t]
    \begin{threeparttable}
    \centering
    \scriptsize
    \caption{
        Comparison of the computational costs and sampling efficiencies at inference time.
        This does not include the time required for generating data and training models.
    }
    \label{tab:runtimes}
    \begin{tabularx}{\linewidth}{@{}lCCCC@{}}
    \toprule
        & \textbf{\# simulations} 
        & \textbf{efficiency}
        & \textbf{CPU hours}   
        & \textbf{wall time} \\
    \midrule
    nautilus & \num{4462500}      & \qty{1.12}{\percent}  & \num{13450} & \qty{8.5}{\day}   \\
    FMPE     & ---                & ---                   & ---\textsuperscript{1}     & \qty{12}{\minute} \\
    FMPE+IS  & \num{384691}       & \qty{13.00}{\percent} & \num{342}   & \qty{3.5}{\hour}\textsuperscript{2}  \\
    NPE      & ---                & ---                   & ---\textsuperscript{1}     & \qty{12}{\second} \\
    NPE+IS   & \num{1999354}      & \qty{2.50}{\percent}  & \num{1777}  & \qty{18.5}{\hour}\textsuperscript{2} \\
    \bottomrule
    \end{tabularx}
    \begin{tablenotes}
        \item \textsuperscript{1}Sampling standard FMPE / NPE uses a GPU.
        \quad \textsuperscript{2}Assuming 96 CPUs (which is what we used for running \texttt{nautilus}); in practice, this is arbitrarily parallelizable.
    \end{tablenotes}
    \end{threeparttable}
\end{table}

\section{Conclusions}

We compared flow matching posterior estimation (FMPE), a new approach to exoplanet atmospheric retrieval based on CNFs, with both neural posterior estimation (NPE) using DNFs and (ML-enhanced) nested sampling as implemented by \texttt{nautilus}.
Both FMPE and NPE yielded good agreement with the reference posterior, while reducing inference times by orders of magnitude compared to nested sampling. 
Notably, FMPE demonstrated slightly higher accuracy and significantly shorter training times than NPE.
Combining both approaches with neural importance sampling, we matched the accuracy of an established nested sampling algorithm, \texttt{nautilus}, while retaining a runtime advantage, in particular assuming an amortized setting.  
One limitation of this ongoing work is that we have only considered a single benchmark retrieval. 
In future work, we will study the properties of FMPE (with and without importance sampling) more systematically.
Our results encourage a broader adoption of SBI approaches for AR to combine high accuracy and diminishing retrieval costs not only in the analysis of real observational data (e.g., from JWST), but also during the design phase of new instruments and missions for exoplanet science, such as HWO \citep{Clery_2023} or LIFE \citep{Quanz_2022}.
\looseness=-1

\section{Acknowledgements}
The authors thank Malavika Vasist and Evert Nasedkin for their help with understanding their data generation code.
Further thanks go to Vincent Stimper for insightful discussions and help with the normflows package.
TDG acknowledges funding from the Max Planck ETH Center for Learning Systems.
MD thanks the Hector Fellow Academy for support.
Part of this work has been carried out within the framework of the National Centre of Competence in Research PlanetS supported by the Swiss National Science Foundation under grants 51NF40\_182901 and 51NF40\_205606. 
DA and SPQ acknowledge the financial support of the SNSF.
DA acknowledges funding through the European Space Agency’s Open Space Innovation Platform (ESA OSIP) program.

This work has benefited from various open-source Python packages, including
corner.py \citep{ForemanMackey_2016},
dynesty \cite{Speagle_2020},
glasflow \citep{Williams_2023},
nautilus \citep{Lange_2023},
normflows \citep{Stimper_2023},
numpy \citep{Harris_2020},
scipy \citep{Virtanen_2020},
spectres \citep{Carnall_2017},
ott-jax \citep{Cuturi_2022},
petitRADTRANS \citep{Molliere_2019},
PyMultiNest \cite{Buchner_2014},
torch \citep{Paszke_2019}, and
torchdiffeq \citep{Chen_2018b}.

\bibliography{main}

\appendix
\section{Appendix}

\begin{figure*}[t]
    \centering
    \includegraphics[scale=1]{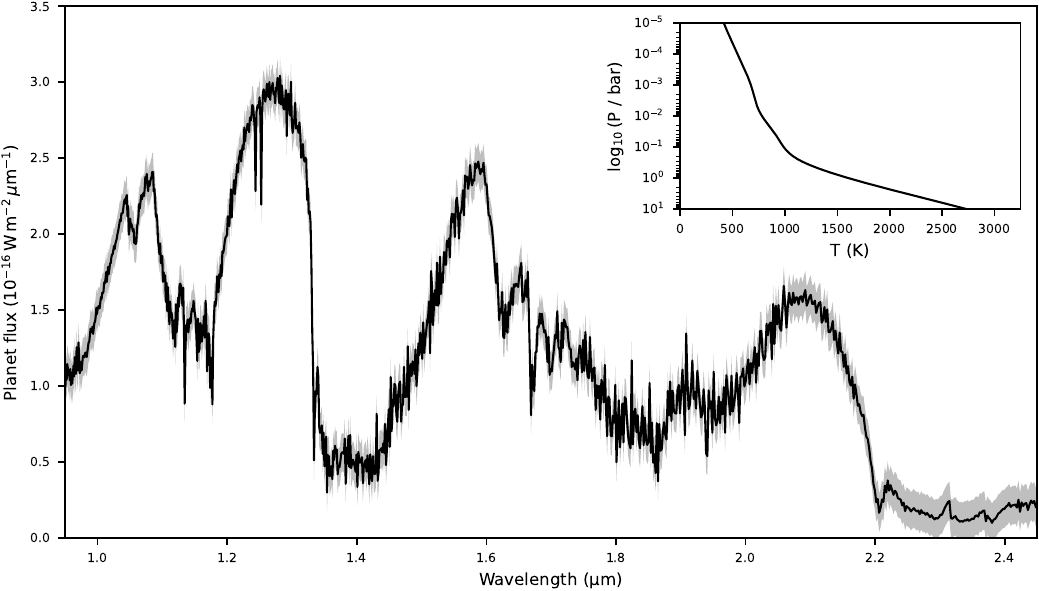}
    \caption{
        Simulated emission spectrum and pressure-temperature profile corresponding to $\theta_0$ (i.e., our benchmark case).
        The shaded area marks the $1\sigma$ error bars we assumed in the likelihood, both for nested sampling and for the training data generation.
    }
    \label{fig:spectrum}
\end{figure*}

\begin{figure*}[t]
    \centering
    \includegraphics[scale=1]{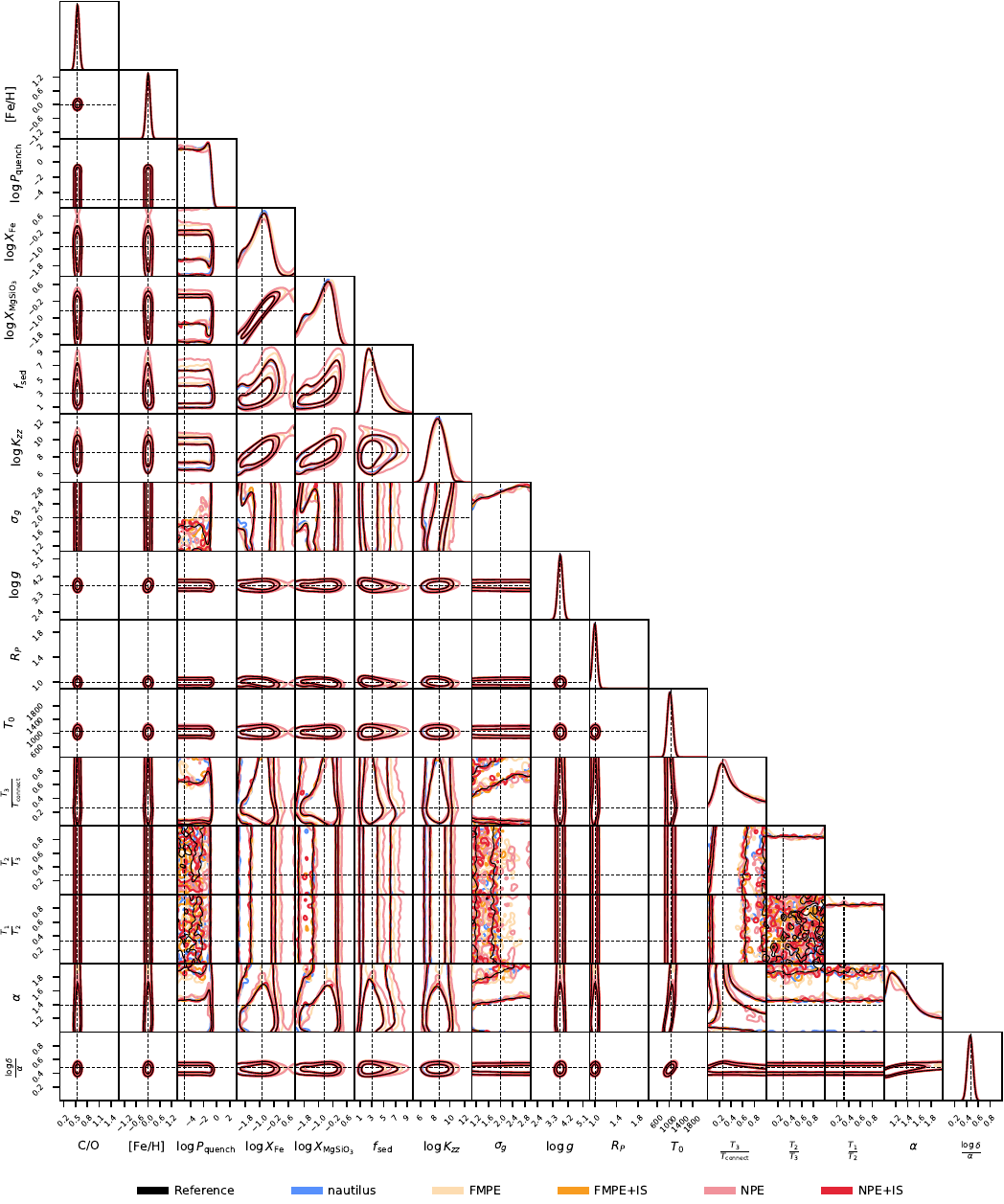}%
    \caption{
        Results of our benchmark atmospheric retrieval:
        This corner plot shows a comparison of the 1D and 2D marginal posterior distributions for the different inference methods (\texttt{nautilus} is our nested sampling baseline).
        The true value $\theta_0$ is marked by the dashed lines.
        The axes limits are set to the ranges of the respective priors.
        This figure is best viewed digitally.
    }
    \label{fig:cornerplot}
\end{figure*}

\begin{figure*}
    \centering
    \includegraphics[]{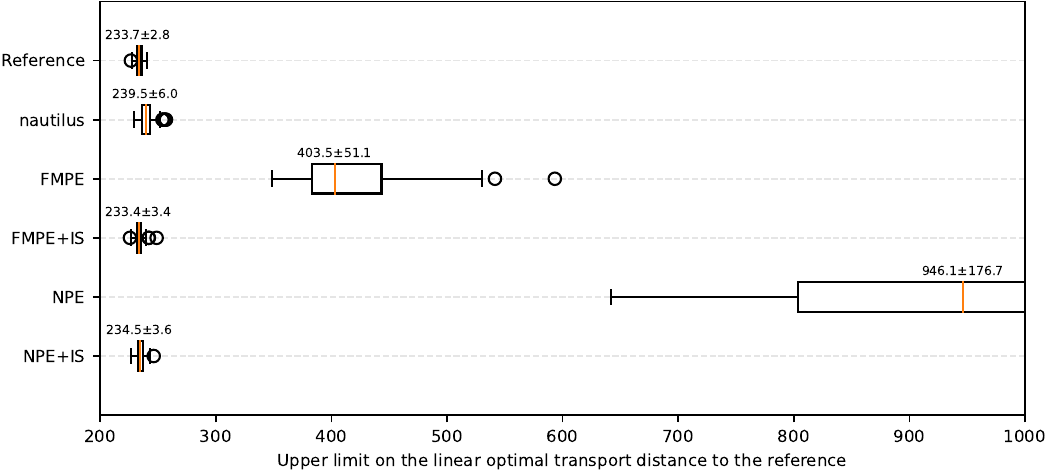}
    \caption{
        Upper bounds on the linear optimal transport distance between our reference posterior and the different estimators:
        We use the \texttt{ott-jax} package to estimate the linear optimal transport (OT) distance between the reference posterior and the different estimates (including the reference itself, to establish the general scale) by treating them as point clouds:
        For each method, we randomly choose 10\,k posterior samples and compare them against an equal-sized random subset of the reference posterior by computing an upper bound on the OT distance using a Sinkhorn solver.
        We repeat this 100 times for each method and compute the median, which we take as an estimate of the distance to the reference.
        The resulting pattern matches the one from \cref{tab:jsd-values}.
    }
    \label{fig:ot-results}
\end{figure*}

This appendix contains \cref{tab:atmospheric-parameters,fig:spectrum,fig:cornerplot,fig:ot-results} that were referenced in the main text.

\begin{table*}[t]
    \begin{threeparttable}
    \centering
    \caption{
        The 16 atmospheric parameters of interest that we consider in this work, including the priors used for the data generation and the parameter values $\theta_0$ of the spectrum used as the benchmark case.
        See also tables 1 and 2 in \citet{Vasist_2023}.
    }
    \label{tab:atmospheric-parameters}
    \begin{tabularx}{\linewidth}{llS[table-format=4.2]X}
    \toprule
    \textbf{Parameter}         & \textbf{Prior}                & \textbf{$\theta_0$ value} & \textbf{Meaning} \\                                                
    \midrule
    ${\rm C/O }$               & $\mathcal{U}(0.1, 1.6)$       &    0.55    & Carbon-to-oxygen ratio \\
    $\left[{\rm Fe/H}\right]$  & $\mathcal{U}(-1.5, 1.5)$      &    0.00    & Metallicity \\
    $\log P_{\rm quench}$      & $\mathcal{U}(-6.0, 3.0)$      &   -5.00    & Pressure at which \ch{CO}, \ch{CH_4} and \ch{H_2O} abundances become vertically constant \\
    $\log X_{\rm Fe}$          & $\mathcal{U}(-2.3, 1.0)$      &   -0.86    & Scaling factor for equilibrium cloud abundances (\ch{Fe}) \\
    $\log X_{\rm MgSiO_3}$     & $\mathcal{U}(-2.3, 1.0)$      &   -0.65    & Scaling factor for equilibrium cloud abundances (\ch{MgSiO_3}) \\
    $f_{\rm sed}$              & $\mathcal{U}(0.0, 10.0)$      &    3.00    & Sedimentation parameter \\
    $\log K_{zz}$              & $\mathcal{U}(5.0, 13.0)$      &    8.50    & Vertical mixing parameter \\
    $\sigma_g$                 & $\mathcal{U}(1.05, 3.0)$      &    2.00    & Width of cloud particle size distribution (log-normal) \\
    $\log g$                   & $\mathcal{U}(2.0, 5.5)$       &    3.75    & (Logarithm of) surface gravity \\
    $R_P$                      & $\mathcal{U}(0.9, 2.0)$       &    1.00    & Planet radius (in Jupiter radii) \\
    $T_0$                      & $\mathcal{U}(300, 2300)$      & 1063.60    & Interior temperature of the planet (in Kelvin) \\
    $T_3 / T_{\rm connect}$    & $\mathcal{U}(0.0, 1.0)$       &    0.26    & \multirow{5}{=}{Parameters that describe the pressure-temperature profile (i.e., temperature as a function of pressure). The forward simulator uses a spline-based version of the parameterization scheme proposed in \citet{Guillot_2010}.} \\
    $T_2 / T_3$                & $\mathcal{U}(0.0, 1.0)$       &    0.29    & \\
    $T_1 / T_2$                & $\mathcal{U}(0.0, 1.0)$       &    0.32    & \\
    $\alpha$                   & $\mathcal{U}(1.0, 2.0)$       &    1.39    & \\
    $\log \delta / \alpha$     & $\mathcal{U}(0.0, 1.0)$       &    0.48    & \\
    \bottomrule
    \end{tabularx}
    \end{threeparttable}
\end{table*}

\end{document}